\documentstyle[prb,aps,epsfig]{revtex}

\begin{document}

\title{Fully Frustrated Ising System on a 3D Simple Cubic Lattice: Revisited}

\author{L.W. Bernardi, K. Hukushima and H. Takayama\\
Institute for Solid State Physics, University of Tokyo\\
Roppongi, 7-22-1 Minato-ku, Tokyo 106-8660, Japan}

\maketitle

\abstract{Using extensive Monte Carlo simulations, we clarify the critical
behaviour of the 3 dimensional simple cubic Ising Fully Frustrated system.
We find two transition temperatures and two long range ordered phases.
Within the present numerical accuracy, the transition at higher temperature is 
found to be second order and we have extracted the standard critical exponent
using finite size scaling method. On the other hand, the transition at lower 
temperature is found to be first order. It is argued that entropy plays a 
major role on determining the low temperature state.}



\section{Introduction}

One of the nowadays trends in statistical physics is the study of disorder. 
The Spin Glass (SG) problem is the most typical subject in the field which, 
despite huge theoretical and experimental works, has not yet been solved. 
Besides disorder Spin Glass involves also frustration which is known to give 
rise to peculiar, non-trivial, behaviors, especially in Ising systems.  
It is therefore considered as a promising approach to first understand 
Fully Frustrated (FF) systems without disorder and then come closer to 
Spin Glass by adding disorder~\cite{kirkpatrick81}. From this point 
of view one of the present authors (L.W.B.) and Campbell investigated the 
critical behavior along the line from the 3 dimensional (3D) FF Ising model to 
the $\pm J$ Ising SG model~\cite{campbell96}. In order to get further insights 
along this line, we have re-investigated the 3D FF Ising system on a 
simple cubic lattice which was once studied more than a decade ago 
\cite{kirkpatrick81,chui82,blank84,grest85,diep85,narita86}
but has been left without thorough understandings. The purpose of the 
present paper is to report our new numerical results on nature of the phase 
transitions in this model.

Chui et al. \cite{chui82} were the first to point out, by means of the 
Bethe Peierls approach,  that the system exhibits a first-order phase 
transition at a finite temperature. They also estimated the degeneracy of 
ground states which grows proportionally to $2^{\frac{L^2}{4}}$ with 
$L$ being the linear dimension of the system. Blanckschtein et 
al~\cite{blank84} (hereafter referred to as BMB) developed a renormalization 
group (RG) analysis ($\epsilon$-expansion) based on the Landau-Ginzburg-Wilson 
(LGW) Hamiltonian which describes critical behavior of the system of interest. 
Their main result is that the transition is a weakly first order one. They 
also obtained explicitly the 16-fold degenerated ordered configurations of 
site magnetization just below the transition temperature. Grest~\cite{grest85}
attempted to prove BMB's results numerically but the result was inconclusive 
since the signature of a first-order transition, i.e., a  bimodal distribution
in the energy  histogram, was not seen. 
 
The first numerical study was done by Kirkpatrick~\cite{kirkpatrick81} which 
showed that this system exhibit a second order phase transition at 
$T_c =1.25J$ while geometrical argument leads to the absence of phase 
transition.  Diep et al~\cite{diep85} (hereafter referred to as DLN) carried 
out the most extensive numerical study on this system and obtained the 
following results. The  transition from the paramagnetic phase to an ordered 
phase occurs at $T_c =1.355J$ for the infinite size limit and is of second 
order. At lower temperatures the system moves to another phase, in which 
disorder is located along unidirectional lines which in turn form a 
periodic array. DLN also mentioned the existence of a crossover behavior at 
a temperature higher than $T_c$.
Narita et al~\cite{narita86} studied the system by looking at configurations 
of wrong (or unsatisfied) bonds, and suggested occurrence of a 
Kosterlitz-Thouless like transition and also note the existence of two 
and possibly three transitions. 

In the present work, by making combined use of the standard Monte Carlo (MC) 
method and the exchange MC method \cite{MCE}, the nature of the critical 
behavior in the 3D FF Ising system is clarified. There exists two ordered 
phases both with long range ordered magnetization patterns. The pattern in 
the higher 
temperature phase below the transition temperature $T_{ c1}$ is one of 
the 16-fold degenerated states predicted by 
the BMB theory, while the one in the lower temperature phase below the 
transition temperature $T_{ c2}$ is one of the 24-fold degenerated states 
found by DLN at lowest temperatures. A mechanism to derive this lower 
temperature phase, which we call the ``$4J$-excitation'', is proposed. 
Within the present numerical accuracy, the transition at $T_{ c1}$ 
($T_{ c2}$) is of second (first) order. The crossover found by DLN above 
$T_{ c1}$ tends to vanish for larger systems. 

The paper is organized as follows. In the next section,  
we will present the model and the methods we have used. Geometrical 
considerations about the ground states 
and the $4J$-excitation are given in 
Sec.~\ref{Gr4J}. In Sec.~\ref{results} the simulated results for various 
quantities, in the whole temperature range, are presented, and in 
Sec.~\ref{discuss} the nature of the two ordered phases and the associated 
transitions are discussed in some details. The last section is devoted to the 
concluding remarks.

\section{Model and Methods}
\label{metm}
\begin{figure}[ht]
\begin{center}
\epsfig{file=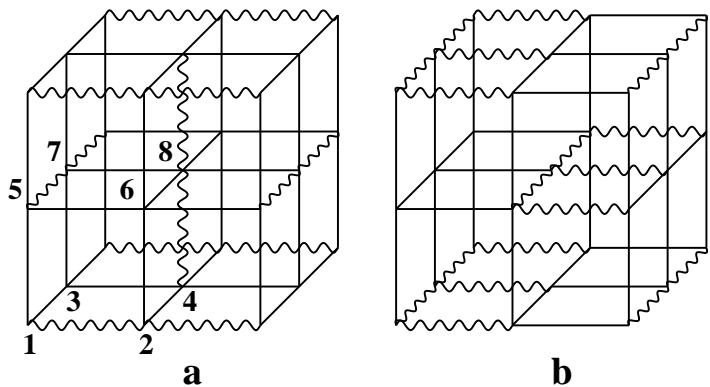,height=5cm}
\end{center}
\caption{\small Structure of the FF 3D SC (a) DLN and (b) Comb-type (BMB) 
model. The (wavy) thin line represent the (anti-)ferromagnetic  bond. The 
numbers labels the sub-lattices used during the simulation.}
\label{FFCS}
\end{figure}

The model studied is described by the standard Hamiltonian,

\begin{equation}
H = - \sum_{<i,j>} J_{ij}S_iS_j,
\label{eq:1}
\end{equation}

\noindent
where $S_i$'s are Ising spins and $J_{ij}$'s are nearest-neighbor
interactions which take value +1 or -1 following patterns described
in Fig.~\ref{FFCS}.a. We call it the DLN model \footnote{Although the first 
study was done by Kirkpatrick, we will rely on the paper by Diep et al. and 
hence the name DLN}. Although its bond pattern is different from the one of 
Fig.~\ref{FFCS}.b (BMB model), the critical behaviour of the two models is 
expected to be the same because every plaquette is frustrated in both 
systems and, in fact, one  can go from one system to the other by gauge 
transformation. However in the present simulation we have examined only the 
DLN model.

In order to investigate critical properties of the model, we have used 
two Monte Carlo (MC) techniques both making use of the heat-bath
updating. One is the standard MC method which has been used to obtain time
evolution, including gradual cooling or heating process, of
quantities of interest. The other is the exchange MC method~\cite{MCE}. The
basic idea of this method is to run simultaneously several replicas
having a common bond configuration but attached to heat baths of
different temperatures, and to exchange stochastically a pair of them
according to the detailed balance condition for the combined system of
all the replicas. The exchange MC (EMC) method is quite efficient to examine 
equilibrium properties of such a system with many degenerate
states.

In the present work the following quantities are evaluated for systems 
with $N=L^3$ spins under the periodic boundary condition:
\begin{itemize}
\item[($a$)] Total internal energy, $E$, and specific heat $C_v$.

\item[($b$)] Overlap distribution defined by

\begin{equation}
P(q)=\sum_{t = 1}^{N_{\rm mcs}}
\delta \left( q-\frac{1}{N}\sum_{i=1}^{N}S_i^a(t)S_i^b(t) \right)
\end{equation}

where $N_{\rm mcs}$ is the number of MC steps, and $a$, $b$ label two
different replicas. 

\item[($c$)] 
Sub-lattice energy, $\epsilon_{\alpha}$, defined by

\begin{equation}
\epsilon_{\alpha} = {8 \over N} \sum_{i \in \alpha}  
\frac{1}{N_{\rm mcs}}\sum_{t = 1}^{N_{\rm mcs}}\sum_j^{\rm n.n.}
-J_{ij}S_i(t)S_j(t),
\end{equation}

where $\alpha$ indicates the sub-lattices ($\alpha = 1,...,8$) which are 
defined by the eight corners of one of the unit cube (see Fig.~\ref{FFCS}.a).

\item[($d$)] Site magnetization defined by

\begin{equation}
M_i^s = \frac{1}{N_{\rm mcs}}\sum_{t = 1}^{N_{\rm mcs}}S_i(t).
\end{equation}

\end{itemize}

\section{Ground States and 4$J$-Excitation}
\label{Gr4J}
\begin{figure}[ht]
\begin{center}
\epsfig{file=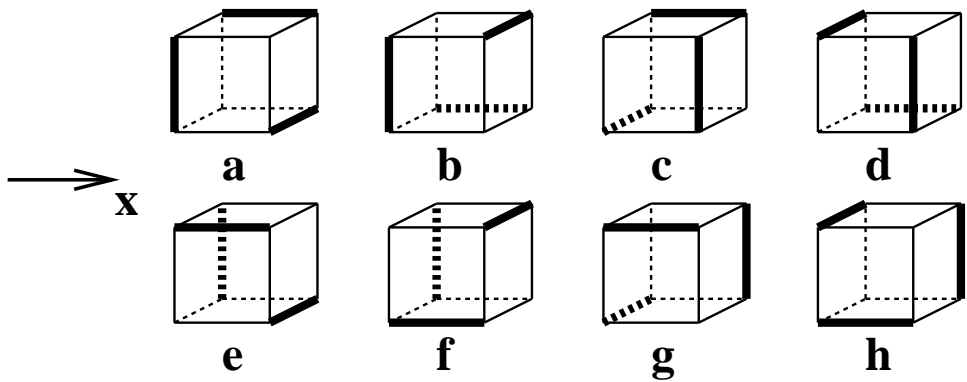,height=5cm}
\end{center}
\caption{\small The eight ways of minimizing the energy of the unit cube 
with three wrong bonds (represented by the bold lines).}
\label{fig:8w}
\end{figure}

A ground state of the Ising system described by eq.~\ref{eq:1} is a spin 
configuration having the minimum number of wrong (or unsatisfied)
bonds which are defined as $-J_{ij}S_iS_j=+J$. Since all plaquettes in 
the present 3D FF model are frustrated, a ground state is obtained
when all the plaquettes have the minimum number of wrong bonds which
is unity. For a ground state of the unit cube, it is easy to see that 
only three wrong bonds are required, thereby none of them are on a
common plaquette and there are two corners which are not touched by
them. Actually there are eight (multiplied by two when global inversion 
of spin is taken into account) ways of minimizing the energy of the 
unit cube as shown in
Fig.~\ref{fig:8w}. The generic way to create a ground state of the total 
system, is to
pile up these cubes with the only constraint that the wrong bond is at the same
place for each common plaquette. For example, the neighbor of $a$ in the 
$x$ direction can be either $c$ or $g$. Let us introduce what we will 
call $periodic$ ground states. If one take the $c$ cube as a neighbor
of $a$, the wrong bond oriented in the $x$ direction
is continuous (in the $g$ case it is not) and the next neighbor of $c$
in the $x$ direction can be $a$. So the pattern can be $acac...$. In this case
the position of the wrong bond perpendicular to the common plaquette has been 
taking into 
account. There are only sixteen $periodic$ ground states since once the first
cube is set there is no choice for the neighboring cube (which gives a 
degeneracy of 8 times 2 for the spin reversal symmetry). In these
periodic ground states only two kinds of sites exist when one looks at the 
sub-lattice energy $\epsilon_{\alpha}$; 6/8 sites with 
$\epsilon_{\alpha}=-2J$ and 2/8 sites with $\epsilon_{\beta}=-6J$.

\begin{figure}[ht]
\begin{center}
\epsfig{file=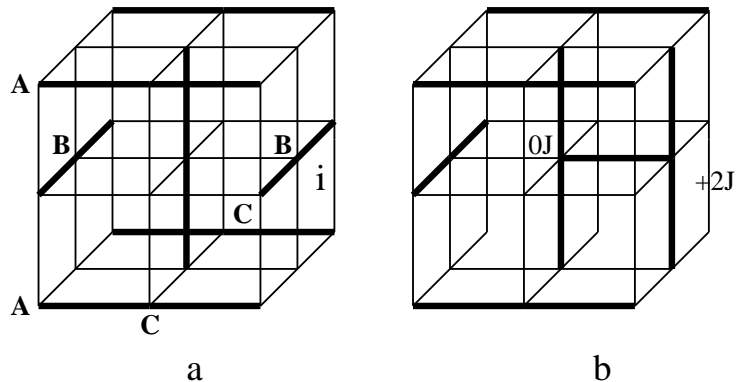,height=5cm}
\end{center}
\caption{\small  ALL the spins of the line AA, BB or CC can be inverted 
without changing the total energy}
\label{fig:s}
\end{figure}

As first pointed out by Kirkpatrick~\cite{kirkpatrick81}, in each of the 
periodic ground states, there are lines of spins
that can be inverted without changing the total energy of the system.
They are the lines AA, BB, and CC in Fig.~\ref{fig:s}.a and were called by 
linear-chain excitations by DLN. Two wrong bonds are
connected to each sites of these lines. The choice of the lines
inverted is restricted to be in the same direction. For example if one
inverts the AA line, one site on the adjacent CC line has only one
wrong bond and thus inversion of this spin (or this CC line) costs
excess energy. Therefore one needs to count how much lines can be
inverted once the direction is chosen. That number is simply given by 
$L^2$/4 and so degeneracy of the ground states is $2^{\frac{L^2}{4}}$, 
which is the same order of magnitude as Chui et al~\cite{chui82} have found.
One of the consequence of this inversion is that the neighboring sites 
of this line which had energy $-2J$ and $-6J$ before the inversion now 
have $-4J$. 

A further interesting fact is that there is an actual process to invert 
all spins on one line which costs an energy of only $4J$. 
In fact if one spin is flipped, for example, the site $i$ on the BB line in 
Fig.~\ref{fig:s}.a,  the wrong-bond pattern changes to the one shown 
in Fig.~\ref{fig:s}b, which has higher energy by $4J$ before the flip. But now 
internal field at the two neighboring sites vanishes so that 
spins on the sites can flip without cost of energy.
This mechanism was called kink-anti-kink pair by 
Kirkpatrick~\cite{kirkpatrick81} but here we call it 
a $4J$-excitation. With the aid 
of one $4J$-excitation all spins on the BB line can be inverted. 
 Strictly speaking, this holds under the periodic boundary condition 
we have adopted. It is important to notice that the presence of one 
$4J$-excitation is associated with entropy of $L\ln2$ which is much larger 
than its energy $4J$. This indicates, and will be demonstrated below that 
entropy plays an important role on determining an ordered phase, if any, at 
lowest temperatures.

\section{Results}
\label{results}
In this section we present the data in the whole temperature range 
simulated. They strongly suggest the existence of two phase transitions in 
the present FF model. The details of their critical nature will be discussed 
in the next section.

\subsection{Specific heat and energy}

\begin{figure}[ht]
\begin{center}
\epsfig{file=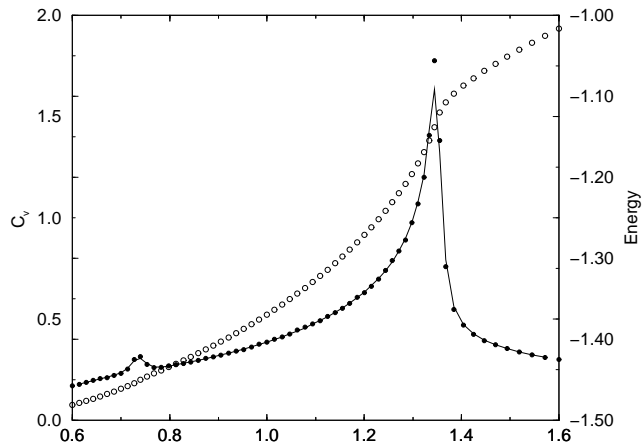,height=6cm}
\end{center}
\caption{\small The ($\bullet$) represent the results of the fluctuation of 
the energy while the solid lines represents the derivative
of the energy. The ($\circ$) represent the energy. L=24}
\label{fig:EFF}
\end{figure}

The results of the specific heat and the energy simulated by the 
exchange MC method are shown in Fig.~\ref{fig:EFF}. They indicate a second 
order phase transition at the temperature $T_{\rm c1} (\simeq 1.35J)$ 
and existence of another transition at $T_{\rm c2} (\simeq 0.70J)$.

\subsection{Overlap distribution}

\begin{figure}[ht]
\begin{center}
\epsfig{file=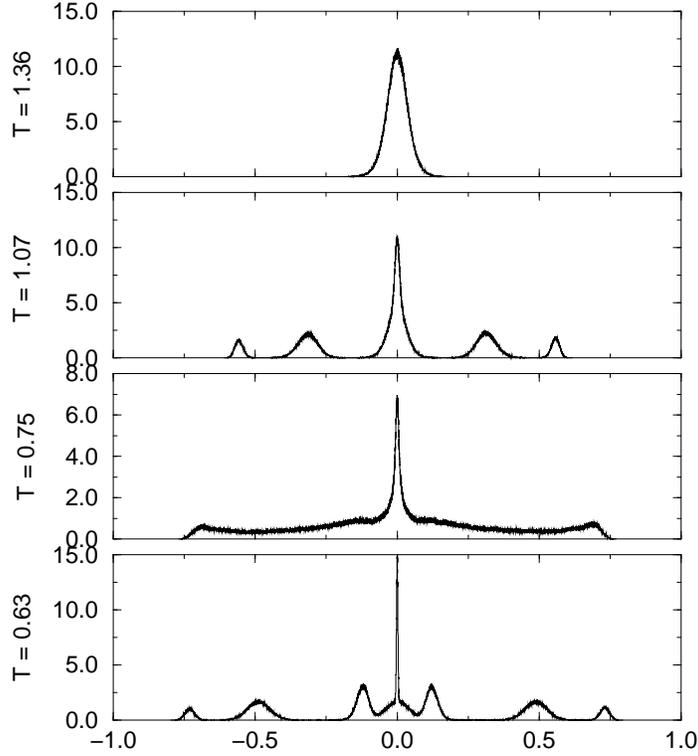,height=10cm}
\end{center}
\caption{\small Overlap Distribution for the FF system. L=24}
\label{fig:PqFF}
\end{figure}

The overlap distribution $P(q)$ simulated by the exchange MC method is shown 
in Fig.~\ref{fig:PqFF}. It exhibits quite different shapes either from 
the one of a simple ferromagnet (with only double peaks at $q=\pm m^2$, 
$m$ being the uniform magnetization) or from the one of mean-field SG model 
(with double peaks at $q=\pm q_{\rm EA}$, $q_{\rm EA}$ being the 
Edwards-Anderson order parameter, and continuous weight between the peaks). 
The characteristic features of $P(q)$ of the present FF model are as 
follows:

\begin{enumerate}
\item It has a single peak centered at $q=0$ in the paramagnetic 
phase ($T > T_{\rm c1}$).
\item At $T_{\rm c1} > T > T_{\rm c2}$, five peaks appear. The peaks 
are narrowed when the size of the system increases. This indicates
that there exist several states which are thermodynamically 
stable. We call the phase of this temperature range the {\it high-$T$ phase}. 
\item At $T=0.75 \simeq T_{\rm c2}$, $P(q)$ exhibits a peak at $q=0$ 
with a continuous distribution in both sides of the peak. This 
implies a certain disordered configuration.
\item Below $T_{\rm c2}$ a seven-peak structure appears, indicating 
occurrence of another ordered state. We call the phase of this temperature 
range the {\it low-$T$ phase}. 
\end{enumerate}

\subsection{Sub-lattice and site quantities}

\begin{figure}[ht]
\begin{center}
\epsfig{file=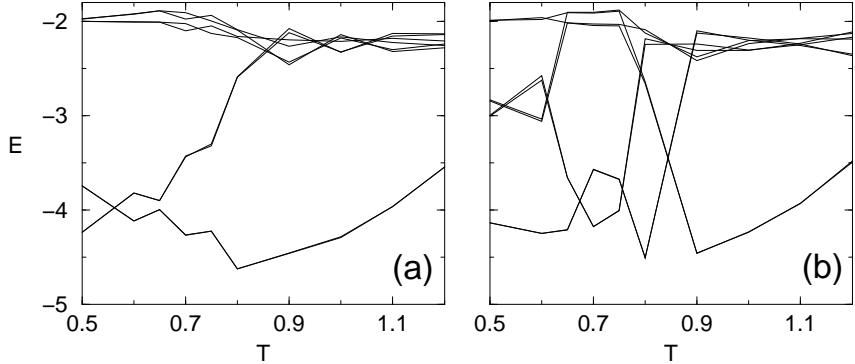,height=5cm}
\end{center}
\caption{ \small Typical sub-lattice Energy. (a) is  obtained  by slow cooling
while (b) is obtain during the heating of a ferromagnetic spin configuration 
from $T=0.50$. All the  data points were  obtained during the same  run for a 
system of size $L=24$}
\label{fig:sublatt}
\end{figure}

We have performed standard MC simulations of gradual cooling or heating 
processes and have looked at evolution of the sub-lattice energy 
$\epsilon_{\alpha}$. The result are shown in Fig.~\ref{fig:sublatt}. The lines 
represent evolution with temperature of the eight $\epsilon_{\alpha}$. For 
each point $10^4$ MCS of annealing were used and the next $10^4$ MCS 
were used to get $\epsilon_{\alpha}$. Each curves in the figure represent 
the result of a single MC run but we have checked that other runs give
qualitatively the same results.

From Fig.~\ref{fig:sublatt}.a we can extract the following characteristic 
features of the gradually cooling process. First of all the sub-lattices are 
linked two by two. In the high-$T$ phase 
($T_{\rm c1} > T > T_{\rm c2}$) there exist two kinds of sub-lattices. 
One with lower energy, $\epsilon_{\alpha}^{(L)}$'s, consists of two 
sub-lattices which are directly checked to be located at opposite corners 
of the cube. The other with higher energies, $\epsilon_{\alpha}^{(H)}$'s, 
consists of the other six sub-lattices. As the system is cooled gradually 
$\epsilon_{\alpha}^{(L)}$'s decrease, while $\epsilon_{\alpha}^{(H)}$'s 
stay nearly constant ($\simeq -2J$). In this phase the system thus 
exhibits a symmetry similar to the one of the periodic ground state 
mentioned previously. In the low-$T$ phase ($T_{\rm c2} > T$) there seem 
to exist three kinds of sub-lattices. Four sub-lattices have 
$\epsilon_{\alpha}$ of $-2J$. The four other sub-lattices are grouped two 
by two with two different $\epsilon_{\alpha}$ which are symmetric around 
$-4J$. This symmetry has already been discussed in the discussion of the 
periodic ground state when the inversion of lines of spin is introduced. 
But this symmetry doesn't hold for case $b$, where the system was heated 
from the ferromagnetic ground state, because the annealing time was 
 not long enough at this low temperature. But it implies that the 
ferromagnetic configuration is not stable at finite temperatures. 

\begin{figure}[ht]
\begin{center}
\epsfig{file=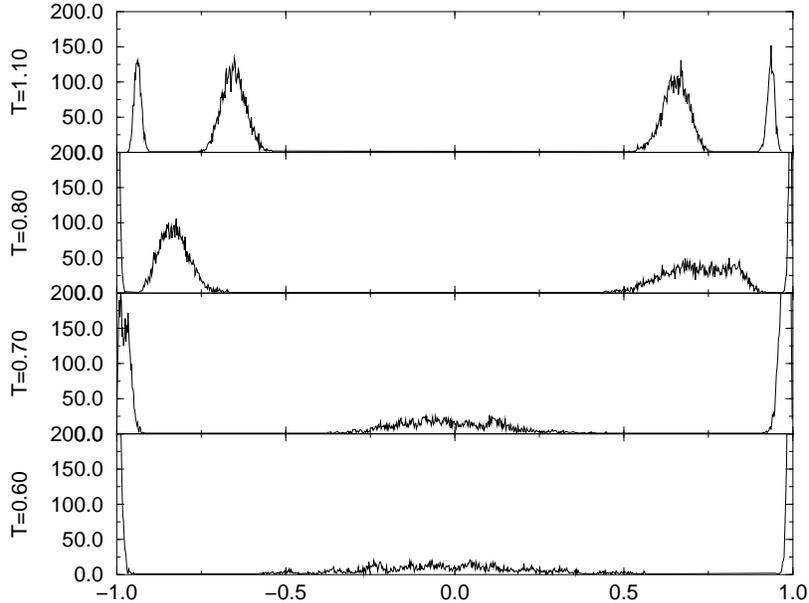,height=8cm}
\end{center}
\caption{\small Histogram of the site-magnetization}
\label{fig:sitemag}
\end{figure}

By the same cooling procedure as above described, we have examined site 
magnetization, $\{ M_i^s \}$. Its distributions in magnitude at four typical 
temperatures are shown in Fig.~\ref{fig:sitemag}. At $T=1.10$, in the high-$T$ 
phase, there are four peaks. However we discuss here about distributions 
against absolute magnitude of $\{ M_i^s \}$, since their signs 
depend on configurations simulated, similarly to those of the sixteen 
periodic ground states. In this sense there are two kinds of $\{ M_i^s \}$,
one with smaller $|M_i|$ and the other with larger one. 
The integration of the peaks centered around 
$|m_1|$ and $|m_2|$ yields 6/8 and 2/8, respectively. Actually, when the 
corresponding spatial pattern of $\{ M_i^s \}$ is visualized, 
it is almost perfectly periodic, and agrees with the one derived by 
the BMB (see Section 5.1). When temperature approaches to $T_{c2}$ the
position of the two peaks tend to saturate to the limiting value, i.e., 
unity. 

At $T=0.70J$ and $0.60J$, in the low-$T$ phase, there is a flat bump around 
zero. The integration of this bump yields weight of 2/8, while the two sharp 
peaks at $\pm 1$ have  weight of 6/8. In the corresponding spatial 
patterns of $\{ M_i^s \}$ we have observed that lines with nearly 
zero magnetization form a periodic lattice, as found by DLN. 
This configuration is attributed to the $4J$-excitation mentioned 
previousely (see Section 5.3).

\section{Discussions}
\label{discuss}

The results of our MC simulations presented in the previous section 
reveal that there exist two ordered phases in the DLN model. In the
following we discuss the nature of the two phase transitions in detail.

\subsection{High-$T$ phase}

When the system is cooled from the paramagnetic phase, an ordered phase 
appears at $T=T_{\rm c1}\simeq 1.35J$ (see Fig.~\ref{fig:EFF}), which we have 
called the high-$T$ phase. It is in good agreement with the one 
theoretically predicted by BMB, which can be easily gauge-transformed 
from their comb-type model to the one used by DLN.

In the DLN model the magnetic unit cell consists of eight unit cubes
shown in Fig.~\ref{FFCS}. In the BMB model, on the other hand, it consists of
sixteen cubes. There exist sixteen degenerated equilibrium
configurations including the spin up-down symmetry. Each elementary cube has 
$|M_i^s|\sim m$ at two diagonally 
opposite sites, and $|M_i^s|\sim m/a$ with $a>1$ at other sites with 
signs specified in such a way that three wrong bonds do not intersect 
and that they do not touch the two sites with larger $|M_i^s|$. We note that 
these configurations are different from the periodic ground states for which 
$a=1$, and that they minimize the LGW Hamiltonian of the present FF 
system~\cite{blank84}.

\begin{figure}[ht]
\begin{center}
\epsfig{file=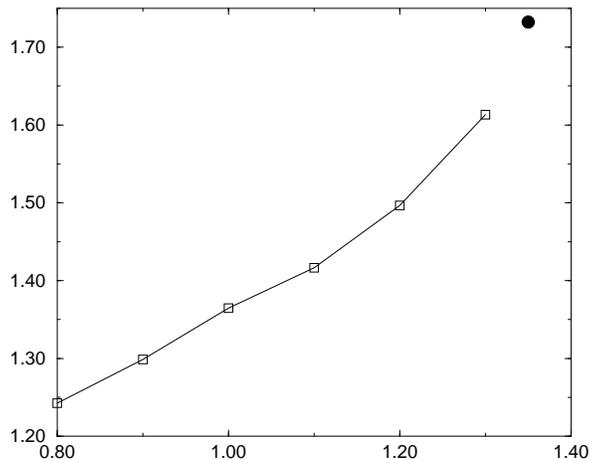,height=6cm}
\end{center}
\caption{\small Ratio of the larger to the smaller site-magnetizations.}
\label{fig:6}
\end{figure}

The ordered configurations mentioned above are quite consistent with
our simulated data described in the previous section. 
Concerned with the overlap distribution $P(q)$ at $T=1.07$ shown in
Fig.~\ref{fig:PqFF}, the positions and weights of the five peaks are what are 
expected  if the system visits the sixteen degenerated configurations with 
equal probability in the exchange MC simulation.
The occurrence of these ordered states is confirmed also by the 
$\{ M_i^s \}$-histogram at $T=1.10$ shown in Fig.~\ref{fig:sitemag}, and by 
the direct inspection on the spatial pattern of $\{ M_i^s \}$ (not shown). 
Furthermore in Fig.~\ref{fig:6} we plot the ratio of positions of the two 
peaks in Fig.~\ref{fig:sitemag} against temperature. The results are 
consistent with the theoretical prediction $a={\sqrt 3}$~\cite{blank84} 
expected to hold at  temperature close to $T_{\rm c1}$.

\begin{figure}[ht]
\begin{center}
\epsfig{file=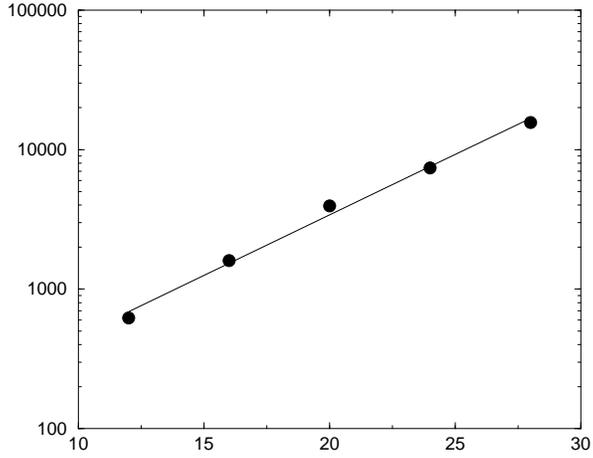,height=6cm}
\end{center}
\caption{\small Time of sub-lattice switching vs the size of the system}
\label{fig:qt}
\end{figure}

In contrast to the previous arguments by Narita et al~\cite{narita86} and 
DLN,  our simulated data strongly suggest that each of the sixteen degenerated 
configurations is long-range ordered (or thermodynamically stable). 
To confirm this we have computed the sub-lattice switching time, 
$\tau_{\rm sc}$. By inspection of time evolution of each sub-lattice 
energy $\epsilon_{\alpha}$ at a fixed temperature close to $T_{\rm c1}$ 
we have observed that it fluctuates between the higher and lower branches 
of $\epsilon_{\alpha}$ in Fig.~\ref{fig:sublatt} with an average interval of 
$\tau_{\rm sc}$ which grows with the size of the system. To extract 
$\tau_{\rm sc}$ more accurately we have used the autocorrelation function 
$q(t)$ which almost saturates to an equilibrium value before 
exhibiting an exponential decay. This exponential decay, at the later stage, 
is attributed to the sub-lattice switching and its relaxation time is 
examined at $T=1.32 \  < T_{\rm c1})$ for sizes $L$ varying from 12 
to 28. The result shown in Fig.~\ref{fig:qt} indicates that $\tau_{\rm sc}$ 
is exponential with an argument growing roughly linearly with $L$. 
This implies that in this 3D FF model the domain-wall free energy is roughly
proportional to $L$ (and not to $L^2$ as for ordinary 3D ferromagnet), 
and that the sub-lattice switching does not occur in the thermodynamic 
limit. 

\begin{figure}[th]
\begin{center}
\epsfig{file=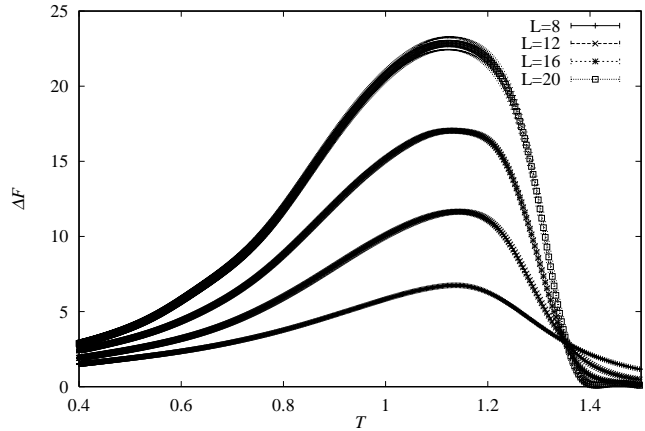,height=6cm}
\end{center}
\caption{\small Results for the defect-free-energy analysis for sizes $L=
8, 12, 16 and 20$}
\label{fig:DFEA}
\end{figure}

The technique usually used to know the domain-wall free-energy, is the
defect-free-energy analysis~\cite{DFEA.ref}. The principle is to introduce
a domain-wall in a system and to see how the energy is affected by this
domain. If the energy increases with the size than the system exhibit 
a phase transition. To obtain the domain wall free energy one compute
\begin{equation}
\Delta F = F_{AP} - F_{P} 
\end{equation}
where $F_{AP}$ and $F_{P}$ are the free energy of the system for the 
anti-periodic (AP) boundaries and periodic (P) boundaries conditions, 
respectively. With standard MC simulations the free energy can be obtained by~:
\begin{equation}
\beta F(T) = \int_{\beta_{min}}^{\beta} d\beta' E(\beta').
\end{equation}
In the simulation the $\beta_{min}$ is set to $1/10J$ and we assume that 
$\Delta F$ is negligible at $\beta_{min}$. The results are shown in 
Figure~\ref{fig:DFEA} where we can see that the maximum of $\Delta F$ also 
have a nearly linear dependence in $L$. We can also note that the two low 
temperature phase have a different slopes for $\Delta F$ against $T$.

\subsection{Critical nature of the transition at $T_{\rm c1}$}

\begin{figure}[ht]
\begin{center}
\epsfig{file=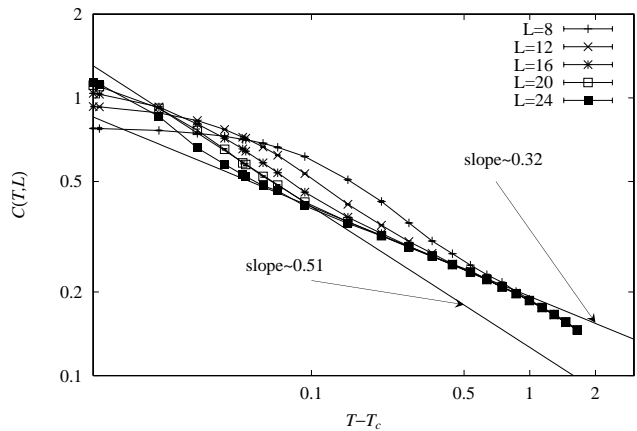,height=6cm}
\end{center}
\caption{\small Size-dependence of the specific heat near $T_{c1}$.}
\label{fig:8}
\end{figure}

According to BMB, the LGW Hamiltonian for describing the phase transition 
in the present FF model is an $n=4$ ``Heisenberg'' model with 
symmetry breaking terms arising from frustration. The RG analysis based 
on the $\epsilon$-expansion method predicts a weakly first order phase 
transition. Within our present numerical accuracy, however, we have not 
detected any evidence which supports a first-order transition at 
$T_{c1}$ (see for example Fig.~\ref{fig:EFF} and further discussions in 
Section 6). The present results are in agreement with the previous numerical 
works by DLN and Grest~\cite{grest85}.

\begin{figure}[ht]
\begin{center}
\epsfig{file=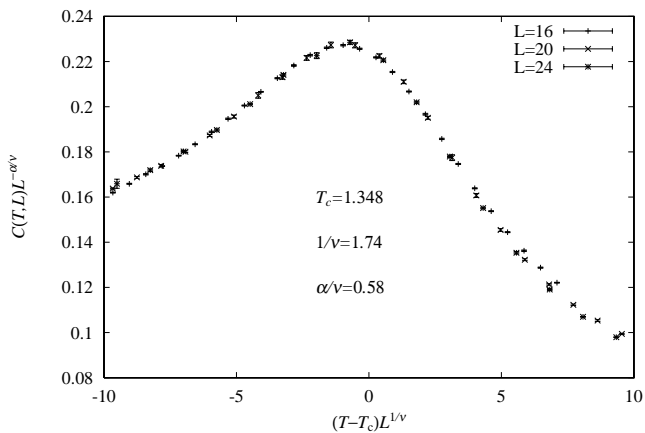,height=6cm}
\end{center}
\caption{Scaling plot of the Specific heat.}
\label{fig:12}
\end{figure}

\begin{figure}[ht]
\begin{center}
\epsfig{file=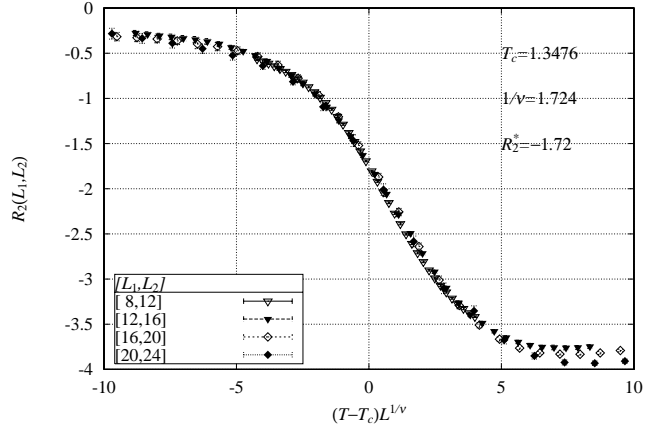,height=6cm}
\end{center}
\caption{Scaling plot of $R_2(L_1, L_2)$.}
\label{fig:13}
\end{figure}

Assuming a second order transition at $T_{c1}=1.347 \pm 0.001$, we have 
obtained the following critical exponents:
\begin{equation}
 \alpha = 0.32\pm 0.02, \ \  
 \nu = 0.56 \pm 0.02, \ \ 
 \beta = 0.25 \pm 0.02, \ \  
 \eta = -0.1 \pm 0.02. 
\end{equation}
These exponents are different from those of the $O(4)$ one and we don't know 
to which class of universality they belongs.
The exponent $\alpha$ of specific heat is estimated either directly as 
shown in Fig.~\ref{fig:8}, or by the finite-size scaling shown in 
Fig.~\ref{fig:12}. 
In relation to our results demonstrated in Fig.~\ref{fig:8}, it is noted 
that the apparent crossover within the paramagnetic phase observed by DLN is 
considered to be an artifact due to finite-size effect. Correspondingly, 
our exponents $\alpha$ and $\nu$ are compatible to those obtained by DLN at 
temperatures above the crossover whose data are considered not affected by 
the finite-size effect. The $\eta$ value given by DLN ($\simeq 0.28$) being 
obtained at $T_{c1}$ is considered to be affected by finite-size effect. 
The Binder cumulant~\cite{binder} method is frequently used to determine the 
exponents $\nu$ and $\beta$. From the later the exponent $\eta$ is calculated 
by the scaling relation $\frac{2\beta}{\nu} =d-2+\eta$. However, for the 
present FF model with the degenerated ordered states below $T_{\rm c1}$ and 
with the presence of the peak at $q=0$ in $P(q)$, a direct application of the 
method is rather complicated. This problem for the Binder cumulant had already
been pointed out for the three states potts models~\cite{potts}.  Instead we 
have analyzed the finite size scaling
of the following ratio~\cite{log_moments}.
\begin{equation}
R_2(L_1,L_2) \equiv \log(\frac{<q^2_{L_2}>}{<q^2_{L_1}>})
/\log(\frac{L_2}{L_1}) = -2{2\beta \over \nu} + g(L_2^{1/\nu}(T-T_c))
\end{equation}
with $g(x)$ being a scaling function and $L_2>L_1 \gg L_2-L_1$. The factor 2 
in  the r.h.s. of the  above equation come out because $<q^2>$ is given by the 
following scaling form:
\begin{eqnarray}
<q^2> & = & < \left ( \frac{1}{N}\sum_{i}^{N}S_i^{\alpha}S_i^{\beta} \right )^2>_T=\frac{1}{N^2}\sum_{i,j}^{N}<S_i^{\alpha}S_j^{\alpha}S_i^{\beta}S_j^{\beta}>_T \\
      & = & \frac{1}{N^2}\sum_{i}^{N}G_{ij}^2 \propto L^{-2(d-2+\eta)} f \left(   L^\frac{1}{\nu}(T-T_c) \right)
\end{eqnarray}
where $<..>_T$ is the thermal average and $f(x)$ another scaling function. Here
$G_{ij}$ is the correlation function $<S_iS_j>_T$ which is given by
\begin{equation}
G_{ij} = G(r=|x_i-x_j|) \propto  r^{-(d-2+\eta)}\exp \left ( \frac{-r}{\xi} \right ) 
\end{equation}
Thus the crossing point of $R_2(L_1, L_2)$ with different set of $(L_1, L_2)$
will be $(T_c,-\frac{4\beta}{\nu})$. The   scaling plot of $R_2(L_1, L_2)$  
giving $T_{c1}$, $\nu$ and $\beta$ is shown in Fig.~\ref{fig:13}. 

\subsection{Low-$T$ phase and nature of the transition at $T=T_{\rm c2}$}

As temperature decreases in the high-$T$ phase, the larger $|M_i^s|$ become 
almost saturated to unity. Then the arguments based on the LGW Hamiltonian, 
in which no restriction on magnitudes of $M_i^s$ is imposed, are expected 
to break down. In fact, as already described in Section~\ref{results}, 
the transition to the low-$T$ phase occurs around $T=T_{\rm c2}\simeq 0.7$. 

In each of spin configurations in the low-$T$ phase realized by gradual 
cooling MC process, one fourth of all the  chains in one of the three 
directions are disordered ($|M_i^s|\simeq 0$), while the others have almost 
saturated 
value ($|M_i^s| \simeq 1$). Furthermore these disordered chains form a 
periodic 2D array. Thus the low-$T$ phase is 24-fold degenerated as pointed 
out by DLN. 
This degeneracy is responsible for the positions and weights of the seven 
peaks structure in $P(q)$ shown in Fig.~\ref{fig:PqFF}.
We argue that the above mentioned order  in the low-$T$ phase is attributed 
to the $4J$-excitation introduced at the end of Section \ref{Gr4J}. As pointed 
out there, the presence of one $4J$-excitation lowers free energy by 
$-TL{\rm \ln}2 + 4J$ as compared with the state without it. This holds true 
so long as the eight chains surrounding the disordered chain (with the 
$4J$-excitation) are firm and  play a role of a {\it cage} of the latter. 
Also we can at least check that two $4J$-excitations on two chains 
perpendicular to each other cannot cross freely. Therefore for the system to 
have maximum free-energy gain a periodic array of $L^2/4$ disordered chains
is realized. 

It is rather hard to check the thermodynamic stability of each of 24 
degenerated states in this phase, not only because MC dynamics 
becomes slower at these low temperatures, but also because system sizes have 
to be large enough for an almost ideal random walk of $4J$-excitations to 
be realized. Also free energy barrier between the different states become 
small as temperature decreases. In fact, as seen in Fig.~\ref{fig:DFEA}, 
magnitudes of defect free energy tend to decrease from around $T_{\rm c2}$ 
with decreasing temperature. But since those with the larger $L$ have still 
the larger values, we expect that each 24 states are thermodynamically 
stable. In this context, it is noted that defect free energy at $T=0$, 
is easily checked to vanish. This, combined with 
existence of the $4J$-excitation, implies that stability of (or free-energy 
barriers between) the ordered states so far discussed, is guaranteed solely
 by the entropy effect.

\begin{figure}[ht]
\begin{center}
\epsfig{file=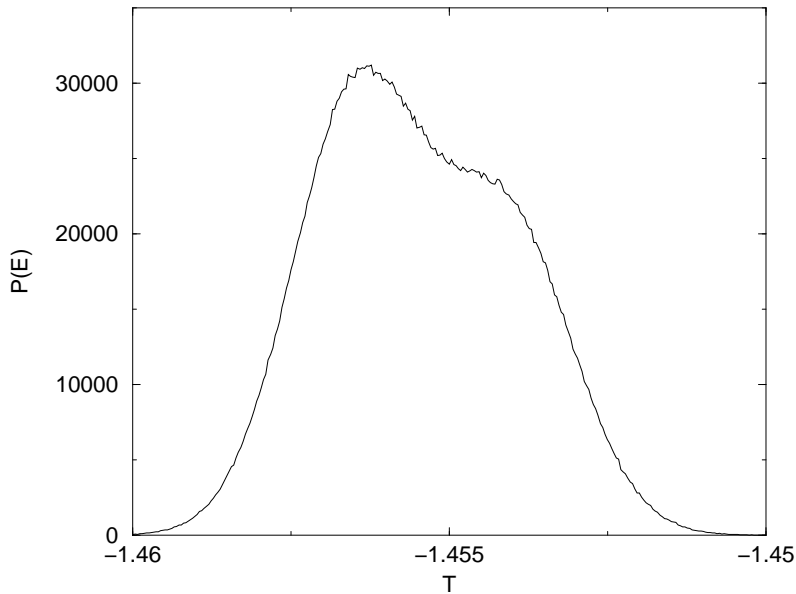,height=8cm}
\end{center}
\caption{\small Histogram of the energy for $L=48$ at $T=0.716$. The 
histogram 
show a double peak structure implying a first order phase transition}
\label{fig:11}
\end{figure}

Lastly we discuss about the nature of the transition at $T_{\rm c2}$ between 
the  high-$T$ and low-$T$ phases. The transition is expected to be of first 
order, since it occurs between spin configurations having different 
symmetries with 16-fold and 24-fold degeneracy. Among our numerical data, 
a peculiar shape of $P(q)$ at $T=0.75$ shown in Fig.~\ref{fig:PqFF}, 
the appearance of 
a significantly large hysteresis in various quantities, and, most plausibly, 
the energy histogram with double peak structure shown in Fig.~\ref{fig:11} 
support this argument. 

\section{Concluding Remarks}

We have shown by extensive MC simulations nature of the phase transitions 
in the 3D FF Ising model. The driving mechanism of the lower transition at 
$T_{\rm c2}$ agrees with what DLN proposed, though we have further introduced 
the ``$4J$-excitation'' mechanism explicitly. We have also claimed that this 
transition is of first order. The nature of the high-$T$ phase below 
$T_{\rm c1}$ is quite consistent with the prediction by the BMB theory. 
Although our results simulated in the present work strongly suggest a 
second-order phase transition at $T_{\rm c1}$, we cannot exclude a 
possibility that it is of first order when studied in systems with larger 
sizes and at closer temperatures to $T_{\rm c}$ as is the case for other 
systems exhibiting a weak first-order transition~\cite{Landau}. The search of 
this possibility is now underway. 

The present work reveals that the transition behavior of the present FF 
model are dominated by a subtle balance of the entropy effects. They are 
expected to be affected sensitively if disorder is added. In particular, 
behavior of the $4J$-excitation in presence of small disorder is of 
interest. The analysis along this direction is also underway.

\section*{Acknowledgements}
We would like to thanks Pr. I.A Campbell for valuable  discussion. Numerical 
calculation were mainly performed on the Fujitsu VPP500 at the Super Computer 
Center
in  the Institute for Solid State Physics. This work is supported by a 
Grant-in-Aids for International Scientific Research Program (n$^0$ 10044064)
and a Grant-in-Aids for Scientific Research Program (n$^0$ 10640362) both from
the Ministry of Education, Science, Sport and Culture. One of 
the authors (L.W.B.) was supported by a Fellowship of the Japanese Society 
for Promoting Science.


\end{document}